\documentclass[twocolumn,aps,prb,superscriptaddress,amsmath,amssymb,superscriptaddress]{revtex4-1}
\newcommand{\pagenumbaa}{1}
\usepackage{graphicx}
\usepackage{color}

\usepackage{verbatim}
\usepackage{amsmath}
\usepackage{float}
\usepackage{hyperref} 
\begin{document}

\title{Narrow optical linewidths and spin pumping on charge-tunable, close-to-surface self-assembled quantum dots in an ultra-thin diode}

\author{Matthias C. L\"obl}
\affiliation{Department of Physics, University of Basel, Klingelbergstrasse 82, CH-4056 Basel, Switzerland}
\author{Immo S\"ollner}
\affiliation{Department of Physics, University of Basel, Klingelbergstrasse 82, CH-4056 Basel, Switzerland}
\author{Alisa Javadi}
\affiliation{\mbox{Niels Bohr Institute, University of Copenhagen, Blegdamsvej 17, Copenhagen DK-2100, Denmark}}
\author{Tommaso Pregnolato}
\affiliation{\mbox{Niels Bohr Institute, University of Copenhagen, Blegdamsvej 17, Copenhagen DK-2100, Denmark}}
\author{R\"udiger Schott}
\affiliation{Lehrstuhl f\"ur Angewandte Festk\"orperphysik, Ruhr-Universit\"at Bochum, D-44780 Bochum, Germany}
\author{Leonardo Midolo}
\affiliation{\mbox{Niels Bohr Institute, University of Copenhagen, Blegdamsvej 17, Copenhagen DK-2100, Denmark}}
\author{Andreas V. Kuhlmann}
\affiliation{Department of Physics, University of Basel, Klingelbergstrasse 82, CH-4056 Basel, Switzerland}
\author{S\o ren Stobbe}
\affiliation{\mbox{Niels Bohr Institute, University of Copenhagen, Blegdamsvej 17, Copenhagen DK-2100, Denmark}}
\author{Andreas D. Wieck}
\affiliation{Lehrstuhl f\"ur Angewandte Festk\"orperphysik, Ruhr-Universit\"at Bochum, D-44780 Bochum, Germany}
\author{Peter Lodahl}
\affiliation{\mbox{Niels Bohr Institute, University of Copenhagen, Blegdamsvej 17, Copenhagen DK-2100, Denmark}}
\author{Arne Ludwig}
\affiliation{Lehrstuhl f\"ur Angewandte Festk\"orperphysik, Ruhr-Universit\"at Bochum, D-44780 Bochum, Germany}
\author{Richard J. Warburton}
\affiliation{Department of Physics, University of Basel, Klingelbergstrasse 82, CH-4056 Basel, Switzerland}

\begin{abstract}
We demonstrate full charge control, narrow optical linewidths, and optical spin pumping on single self-assembled InGaAs quantum dots embedded in a $162.5\,\text{nm}$ thin diode structure. The quantum dots are just $88\,\text{nm}$ from the top GaAs surface. We design and realize a p-i-n-i-n diode that allows single-electron charging of the quantum dots at close-to-zero applied bias. In operation, the current flow through the device is extremely small resulting in low noise. In resonance fluorescence, we measure optical linewidths below $2\,\mu\text{eV}$, just a factor of two above the transform limit. Clear optical spin pumping is observed in a magnetic field of $0.5\,\text{T}$ in the Faraday geometry. We present this design as ideal for securing the advantages of self-assembled quantum dots -- highly coherent single photon generation, ultra-fast optical spin manipulation -- in the thin diodes required in quantum nano-photonics and nano-phononics applications.
\end{abstract}

\maketitle

\setcounter{page}{\pagenumbaa}
\thispagestyle{plain}

\section{Introduction}
\label{sec:introduction}
Single self-assembled quantum dots are a source of high-quality single photons; they are also hosts for single spins \cite{Kroutvar2004,Press2008,Greilich2007,Lu2010,Yifmmode2010,Greilich2009,Prechtel2016}. Their large optical dipole moment enables fast initialization, manipulation, and readout of spin states all by optical means \cite{Gerardot2008,Press2008,Greilich2009,Vamivakas2009,Warburton2013}. In the best case, transform-limited single photon emission from single quantum dots has been demonstrated \cite{Kuhlmann2014}. These properties are extremely sensitive to the quantum dot environment. The electrical environment can be controlled by embedding the quantum dots in diode heterostructures. This locks the Fermi energy and provides electrical control of the quantum dot charge state. Some of the best performances have been achieved in heterostructures that are $\sim 500\,\text{nm}$ thick with the quantum dot positioned $\sim 300\,\text{nm}$ from the GaAs-air interface \cite{Kuhlmann2013,Kuhlmann2014}. 

The solid-state character of these emitters allows their optical \cite{Lodahl2015} and mechanical \cite{WilsonR2004,Sollner2016} properties to be engineered by nano-structuring. For instance, embedding emitters in a membrane leads to the suppression of out-of-plane radiation modes through total internal reflection; control of the in-plane modes can be achieved via lateral patterning of the membrane. Cavities and waveguides can be engineered by creating defects in a photonic crystal bandgap structure. Single photons can be routed on chip, and controlled by single two-level systems \cite{Javadi2015}. Likewise, engineering the mechanical properties can create phononic structures with the aim of controlling the quantum-dot--phonon interaction \cite{WilsonR2004,Sollner2016}. In all these applications, the basic building block is a thin GaAs membrane. It is crucial that the quantum dots in these thin structures exhibit the same excellent properties of quantum dots in bulk-like structures. This has not been achieved so far.

Typical photonic crystal membranes, in the wavelength regime relevant for InGaAs quantum dots, range in total thickness from $120\,\text{nm}$ to $200\,\text{nm}$ \cite{Reese2001,Pinotsi2011}. The first demonstrations of charge control on quantum dots in photonic crystals used thin p-i-n diode structures \cite{Pinotsi2011,Vora2015}. However, the large built-in electric field in combination with the small thickness of these devices led to a large potential at the position of the quantum dots shifting the Coulomb plateaus to large forward bias voltages. This resulted in high tunneling currents in p-i-n-membrane devices, a possible explanation for the absence of spin pumping in embedded quantum dots \cite{Pinotsi2011}. The quantum dot optical linewidths were relatively high in these structures.

In order to avoid the problems associated with high tunneling currents, we present here a quantum dot diode which operates close to zero bias. The main idea is to incorporate an n-layer within a p-i-n device, resulting in a p-i-n-i-n diode. The intermediate n-layer is fully ionized. Most of the built-in field between the outer p- and n- gates drops between the top p-gate and the intermediate n-layer. The electric field at the location of the quantum dots is therefore much smaller than in a p-i-n diode with equal dimensions. This allows single-electron charging to occur close to zero bias. The p-i-n-i-n diode is used in silicon transistor technology \cite{Nagavarapu_PNPN,Hosseini2015}, albeit with lateral rather than vertical control of the doping. It has also been employed in self-assembled quantum dot devices\cite{Lagoudakis2013,Vora2015} but in these experiments narrow optical linewidths in combination with good spin properties were not achieved. 

We present here a careful design which fulfills a list of criteria. The design rests on a full quantitative analysis of the band bending. It is realized using state-of-the-art GaAs heterostructures \cite{Kuhlmann2014,Ludwig2017}. We present resonant laser spectroscopy on single quantum dots in a $162.5\,\text{nm}$ thick p-i-n-i-n diode with a quantum dot to surface distance of just $88\,\text{nm}$. Deterministic charge control at low bias, narrow optical linewidths, as well as optical spin pumping is demonstrated for these close-to-surface quantum dots. The developed heterostructure is ideal for electrical control of quantum dots in nanostructured membranes for photonic and phononic applications.

\section{The p-i-n-i-n quantum dot heterostructure}
\label{sec:sample}
In the design of this structure we had to fulfill a number of constraints. First, the quantum dots should operate in the Coulomb blockade regime. This allows individual quantum dots to be loaded deterministically with single electrons. Within a Coulomb blockade plateau, the external bias allows some fine tuning of the optical transition frequencies via the dc Stark effect. Second, the dc current flowing through the device should be as small as possible to avoid decoherence processes. This can be guaranteed only if the charging voltage is close to zero bias. Third, the optical linewidths on driving the quantum dot resonantly should be small, close to the transform limit. This places stringent conditions on the level of charge noise that can be tolerated. Fourth, the membrane should have as little optical absorption as possible. Fifth, the membrane should be thinner than $\sim250\,\text{nm}$ to ensure single-mode behavior in waveguide structures. In fact, the fabrication of such nanostructures with a soft-mask technique sets a slightly stronger constraint: $180\,\text{nm}$ is the maximum membrane thickness which can be processed with vertical sidewalls \cite{Midolo2015}. Sixth, the quantum dots must be located close to the center of the diode structure to maximize the coupling to TE-like photonic modes \cite{Lodahl2015}. Seventh, the spin relaxation time should be large so that the spin can be initialized and manipulated. In a diode device at low temperature, this means that the co-tunneling rate between a quantum dot electron spin and the Fermi sea should be suppressed by using a relatively large tunnel barrier \cite{Smith2005}. In such a situation, the spin can be initialized into one of its eigenstates by optical pumping \cite{Atature2006,Dreiser2008,Gerardot2008,Lu2010}. 

\begin{figure}[t!]
\includegraphics[width=0.9\columnwidth]{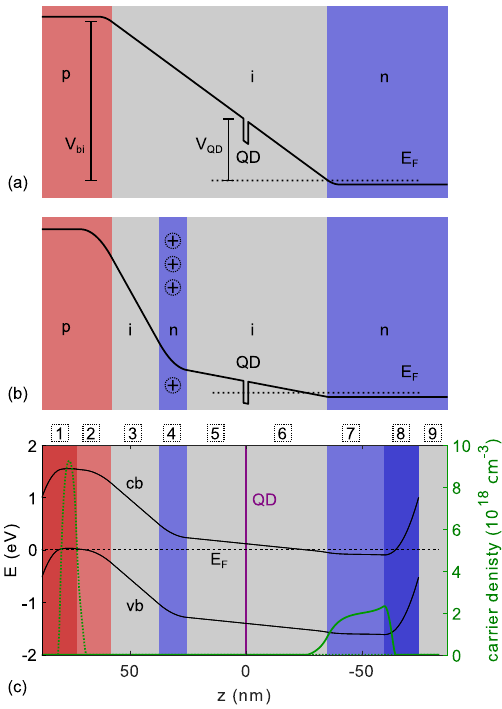}
\caption{(a,b) Schematic conduction band profile of a p-i-n and p-i-n-i-n diode at zero bias voltage. In the p-i-n-i-n structure, an intermediate, fully ionized n-type layer causes band bending, reducing the potential difference between quantum dots and back-gate. In this way, the quantum dots can be charged at a bias voltage close to zero. In contrast, a large positive voltage must be applied to the p-i-n diode. (c) Heterostructure of the investigated samples. Conduction (cb) and valence (vb) band edges are plotted in black and the density of free carriers is plotted in green (dotted line for holes, solid line for electrons). The dashed black line indicates the Fermi level, $E_{\text{F}}$. The purple layer indicates the location of the quantum dots at the center of the membrane. The quantum dots are not included in the band-structure simulation. The diode structure is grown on top of a $1371\,\text{nm}$ thick $\text{Al}_{0.75}\text{Ga}_{0.25}\text{As}$ sacrificial layer enabling selective under-etching. The quantum dots are a distance of $35\,\text{nm}$ away from a back gate consisting of two n-type layers (light and dark blue). The top gate consists of two p-type layers with different doping concentrations (indicated in red). An additional n-type layer is located between quantum dots and top gate. The full heterostructure is constructed as follows: $12.5\,\text{nm}$ intrinsic GaAs (layer 9), $15\,\text{nm}$ n-type GaAs with a doping concentration of $n_{D+}=8.0\cdot 10^{18}\,\text{cm}^{-3}$ (layer 8),  $24.5\,\text{nm}$ n-type GaAs with $n_{D}=2.0\cdot 10^{18}\,\text{cm}^{-3}$ (layer 7), $35\,\text{nm}$ intrinsic GaAs (layer 6), a layer of InGaAs quantum dots, additional $25.5\,\text{nm}$ intrinsic GaAs (layer 5), $12\,\text{nm}$ n-type GaAs with $n_{d}=2.0\cdot 10^{18}\,\text{cm}^{-3}$ (layer 4), $20.5\,\text{nm}$ intrinsic GaAs (layer 3), $15\,\text{nm}$ p-type GaAs with $n_{A}=2.0\cdot 10^{18}\,\text{cm}^{-3}$ (layer 2), $15\,\text{nm}$ p-type GaAs with $n_{A+}=1.0\cdot 10^{19}\,\text{cm}^{-3}$ (layer 1).}
\label{fig:design}
\end{figure}

Fulfilling these constraints is very challenging. It is clearly necessary to work with epitaxial gates, n- and p-type regions in the device, as a metallic Schottky barrier is highly absorbing. A thin p-i-n diode is in principle a possibility. However, at zero bias, there is a very large built-in electric field (Fig.\ \ref{fig:design}(a)). Furthermore, the quantum dots must be positioned at least $30-35\,\text{nm}$ away from the n-type back contact in order to suppress co-tunneling sufficiently. The combination of both constraints means that the quantum dot charges with a single electron only at a large and positive bias, around $\sim 1$ V. Current flow through the device is inevitable under these conditions and high currents are hard to avoid \cite{Pinotsi2011,Carter2013}. The quantum dots could be located closer to the back gate while suppressing co-tunneling by using AlGaAs tunneling barriers. Highly opaque AlGaAs tunneling barriers have been successfully used \cite{Patel2010,Houel2012,Kurzmann2016}. More transparent AlGaAs tunneling barriers require extremely precise control of thickness and Al-content, hard to achieve in practice.

An alternative to the p-i-n diode is a diode with an additional n-layer in the intrinsic region, a p-i-n-i-n device (Fig.\ \ref{fig:design}(b)). The additional n-layer lies in the depletion region of the surrounding p-i-n diode. It is fully depleted such that it becomes positively charged. At zero bias, the total potential drop between p- and n-layers is the same as in the p-i-n diode, but now there is a large drop between the top p-contact and the intermediate n-layer, followed by a small drop between the intermediate n-layer and the back contact. By choosing the location and doping levels of the intermediate n-layer, the device can be designed so that the quantum dot charging voltage lies close to zero volts. 

The p-i-n-i-n design allows in principle all seven criteria to be met. The design is compatible with a $35\,\text{nm}$ i-GaAs tunneling barrier which is known to result in clear Coulomb blockade yet suppresses co-tunneling sufficiently so that spin initialization can be carried out with high fidelity with optical pumping even in the Faraday geometry \citep{Atature2006,Dreiser2008}. The device can be operated close to zero bias, resulting in very small currents. Absorption is minimized by using epitaxial gates instead of metal Schottky gates. The intermediate n-layer is fully ionized and therefore should not result in any additional losses. The entire heterostructure (see Fig.\ \ref{fig:design}(c)) can be made as thin as $176\,\text{nm}$ with the quantum dots located in the center. 

In practice, the performance of a p-i-n-i-n device needs to be tested experimentally. A particular challenge is to achieve narrow optical linewidths for quantum dots just $80-90\,\text{nm}$ away from the free surface as it is a known source of charge noise. By using careful design and state-of-the-art material, we report here success in this endeavor. 

\section{p-i-n-i-n design and device fabrication}
A p-i-n-i-n heterostructure is designed to fulfill the seven criteria. Charges and electric fields are calculated by solving the Poisson equation, either numerically (nextnano) or analytically within the depletion approximation (see \hyperref[sec:appendix a]{appendix A}). In the numerical simulation, the effect of surface depletion due to surface Fermi pinning is taken into account by using a Schottky barrier height of $1\,\text{eV}$ at the surface of the structure. The two approaches give results which are in good quantitative agreement. The calculated band bending and exact layer sequence are shown in Fig.\ \ref{fig:design}(c).

The sample is grown by molecular beam epitaxy. The diode itself is grown on top of a $1371\,\text{nm}$ thick $\text{Al}_{0.75}\text{Ga}_{0.25}\text{As}$ sacrificial layer which enables fabrication of free standing membranes via selective wet etching \cite{Midolo2015}. The first part of the active layer is a $12.5\,\text{nm}$ thick layer of intrinsic GaAs (no.\ 9 in Fig.\ \ref{fig:design}(c)), followed by a back gate consisting of two layers of n-type (silicon-doped) GaAs. The first layer (no.\ 8) is $15\,\text{nm}$ thick and has a high doping concentration $n_{D+}$; the second layer (no.\ 7) is $24.4\,\text{nm}$ thick with a lower doping concentration $n_D$ (see Fig.\ \ref{fig:design}(c) for precise values). A tunnel barrier (no.\ 6) of $35\,\text{nm}$ intrinsic GaAs separates the back gate from a layer of InGaAs quantum dots. Above the quantum dot layer, a $25.5\,\text{nm}$ thick capping layer (no.\ 5) of intrinsic GaAs is grown; subsequently, the intermediate n-type layer (no.\ 4) with a doping density of $n_d$ and a thickness of $12\,\text{nm}$ is grown. Finally, there is a $20.5\,\text{nm}$ layer (no.\ 3) of intrinsic GaAs and a top gate consisting of two $15\,\text{nm}$ thick p-type (carbon-doped) GaAs layers (no.\ 1, 2). The first p-type layer (no.\ 2) has a lower doping concentration ($n_A$) than the second one ($n_{A+}$) (see Fig.\ \ref{fig:design}(c) for precise values). The intention of the very highly doped p-type layer on top of the device is to prevent surface depletion of the top gate and to allow for fabrication of high-quality ohimc p-contacts.

To fabricate devices from the wafer material, first a mesa structure is defined by means of optical lithography. The top gate is etched away around this mesa so that an independent contact to the back gate can be made. A wet chemical process with a diluted mixture of sulfuric acid and hydrogen peroxide ($1\,\text{H}_2\text{SO}_4\,:\,1\,\text{H}_2\text{O}_2\,:\,50\,\text{H}_2\text{O}$) was used for the etching. Subsequently, a contact pad of Au/Ge/Ni is evaporated onto the new etched surface and then annealed at $420\,^{\circ}\text{C}$ resulting in an ohmic contact to the back gate \cite{Goktas2008}. In the next step, we evaporate a pad of $3\,\text{nm}$ titanium followed by $7\,\text{nm}$ of gold on a small part ($\sim 1\,\text{mm}^2$) of the top gate using a shadow mask. On account of the small distance between the top- and the back gate in this device, standard bonding processes were avoided as a precautionary measure. Instead the electrical contacts to the gates were made by affixing the wires to the bond pads with silver paint.

\section{Photoluminescence and resonance fluorescence}
\label{sec:measurements}
The samples were measured in a helium bath cryostat at $4.2\,\text{K}$. Optical experiments were performed with a confocal dark-field microscope with a spot size close to the diffraction limit \citep{Kuhlmann2013a}. All measurements were carried out on two samples processed from the same wafer, denoted as sample 1 and 2 in the following. Both samples fulfill all the requirements that we defined at the outset: a diode-like IV-characteristic with low tunneling currents at small bias voltages, exciton charging transitions at small bias voltages, narrow linewidths in resonance fluorescence, and optical spin pumping.

\begin{figure}[t!]
\includegraphics[width=0.9\columnwidth]{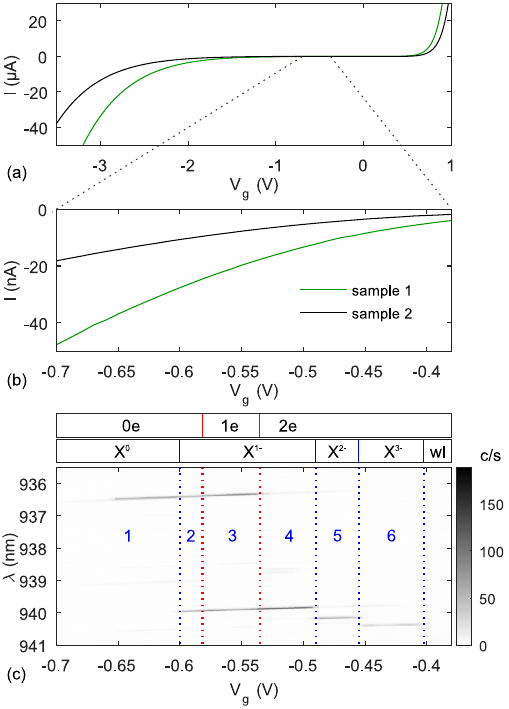}
\caption{(a) IV-curve of two separate samples. Both IV-curves were measured at $4.2\,\text{K}$ and show a typical diode behavior. (b) IV-curve in the voltage regime where excitons of single quantum dots are measured. (c) Photoluminescence (PL) for weak non-resonant excitation ($830\,\text{nm}$) as a function of applied bias voltage for a quantum dot in sample 1. The emission of neutral (X\textsuperscript{0}) and the negatively charged excitons (X\textsuperscript{1-}, X\textsuperscript{2-}, and X\textsuperscript{3-}) is observed. All excitons appear at a low bias voltage where the tunneling current is only several tens of nA. The dotted blue lines indicate the regimes in which the different exciton states become energetically favourable. The dotted red lines indicate the single-electron regime of the quantum dot as measured by resonance fluorescence (RF). Owing to the weak excitation power in PL, the single-electron regime observed in RF coincides with the PL measurement. For high-power non-resonant excitation, the charging steps in the PL can be shifted by optically created space charge.}
\label{fig:PL}
\end{figure}
Plotted in Fig.\ \ref{fig:PL}(a) are the IV-curves of the two samples, both showing diode-like behavior with very low tunneling currents for a large region around $0\,\text{V}$. This excellent electrical behavior is a consequence of both the high material quality of our wafers and the careful contacting of the p-gate. 

We characterize the charging behavior of a single quantum dot by measuring its photoluminescence (PL) as a function of an external bias voltage. Excitation is carried out with a continuous-wave laser with a wavelength of $830\,\text{nm}$ (wetting layer excitation). The voltage applied between top and back gates of the sample changes the energy difference between the back gate Fermi level and the discrete energy levels of the quantum dot. The PL shows clear Coulomb blockade with a series of plateaus, see Fig.\ \ref{fig:PL}(c). We assign these plateaus to the neutral exciton X$\textsuperscript{0}$ and the charged excitons {X\textsuperscript{1-}}, {X\textsuperscript{2-}}, and {X\textsuperscript{3-}}. All charge plateaus appear in reverse bias, in a range between $-0.7\,\text{V}$ and $-0.4\,\text{V}$. At these bias voltages, the tunneling current through the sample is limited to only a few tens of nA for a mesa size of $\sim 15\,\text{mm}^2$ (see Fig.\ \ref{fig:PL}(b)), corresponding to a current density of less than $\sim 3\,\text{nA/mm}^2$.

Our PL-measurements can be interpreted in a majority-minority carrier picture: the optical excitation creates the minority carrier, the hole; the back gate provides majority carriers, electrons. For a $25\,\text{nm}$ tunnel barrier (e.g.\ used in Ref. \cite{Warburton2000,Finley2001,Smith2005,Kuhlmann2013}), electron tunneling is typically much faster than recombination such that once a hole is captured, fast tunneling enables the exciton with the smallest energy to be formed before recombination occurs \cite{Warburton2013}. Abrupt changes in the PL spectrum as a function of bias result. In this work, the tunnel barrier is larger, $35\,\text{nm}$, and interpretation of the PL spectrum is slightly more involved.

In the first region of Fig.\ \ref{fig:PL}(c), the ground state is an empty quantum dot and the lowest energy excited state is X$\textsuperscript{0}$. When a single hole is captured by the quantum dot, it becomes energetically favorable for a single electron to tunnel into the quantum dot, forming an exciton and via recombination a photon at the X$\textsuperscript{0}$ wavelength. 

The first dashed line between regions 1 and 2 in Fig.\ \ref{fig:PL}(c) marks the point at which the X$\textsuperscript{1-}$ and X$\textsuperscript{0}$ energies cross, while the empty quantum dot remains the ground state of the system. In region 2, electrons begin to tunnel into the quantum dot once it has captured a single hole and the X$\textsuperscript{1-}$ line appears. The fact that the X$\textsuperscript{0}$ remains bright at this point, although not as bright as X$\textsuperscript{1-}$, indicates that the electron tunneling time into the quantum dot is comparable to the X$\textsuperscript{0}$ radiative lifetime: recombination can occur before tunneling has created the exciton with lowest energy. We note that the tunneling rate is large enough that no quenching of the resonance fluorescence of $\text{X}^{1-}$ due to an Auger process (by which an electron-hole pair in the X$\textsuperscript{1-}$ decays by ejecting the second electron out of the quantum dot) is expected. The Auger process was demonstrated for thicker tunnel barriers \cite{Kurzmann2016}.

Initially it is perhaps surprising that the X$\textsuperscript{0}$ brightness increases in the regime where the quantum dot ground state is the single-electron state (region 3 of Fig.\ \ref{fig:PL}(c)). These measurements are carried out in the weak excitation regime where hole capture is significantly slower than exciton recombination. The single-electron ground state implies that X$\textsuperscript{0}$ recombination can take place as soon as a hole is captured. We speculate that the presence of an electron in the quantum dot increases the hole capture rate.
 
In the fourth region, the quantum dot is charged with two electrons in its ground state. Thus, capture of a single hole enables the X$\textsuperscript{1-}$ recombination. In this region the intensity of X$\textsuperscript{0}$ is small. X$\textsuperscript{1-}$ recombination leaves behind a single electron. If a hole is captured before tunneling takes place, X$\textsuperscript{0}$ emission is possible. However, this is unlikely with weak optical excitation (the case here) as electron tunneling is faster than hole capture.

Finally, in regions 5 and 6  the energetically favorable excitons are the X$\textsuperscript{2-}$ and X$\textsuperscript{3-}$ states. These states contain one and two electrons in the quantum-dot p-shell, respectively. The tunneling barrier is more transparent for the p-shell than for the s-shell on account of the higher p-shell energy leading to faster tunneling times and therefore less overlap between the plateaus measured in PL. 

The PL experiment establishes that the transition between the 0 and 1e ground states takes place at $-0.6\,\text{V}$, not exactly at the design value of zero. This can be explained by a slight inaccuracy in the doping concentration of the intermediate n-type layer (see \hyperref[sec:appendix b]{appendix B}).

\begin{figure}[t!]
\includegraphics[width=0.9\columnwidth]{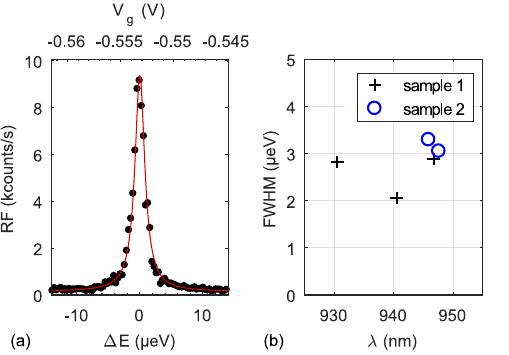}
\caption{(a) Resonance fluorescence of the singly charged exciton X$^{1-}$ measured on the quantum dot shown in Fig.\ \ref{fig:PL}(b). The linewidth obtained by fitting a Lorentzian profile (red line) to the data (black circles) is $1.9\,{\mu}\text{eV}$ FWHM. The count rate is $22.5\,\%$ of the saturation count rate. (b) Average linewidths across the singly charged exciton plateau for five quantum dots in two separate samples. The linewidths lie reproducibly in the range $2-3.5\,{\mu}\text{eV}$.}
\label{fig:RF}
\end{figure}

We turn now to resonant excitation of single quantum dots: this measures the exact optical linewidth. A resonance fluorescence (RF) measurement of the quantum dot presented in Fig.\ \ref{fig:PL}(b) is shown in Fig.\ \ref{fig:RF}(a). The resonant excitation is carried out with a coherent, continuous-wave laser and the reflected laser light is suppressed with a cross-polarized detection scheme \cite{Kuhlmann2013a}. We make use of the Stark shift to sweep the quantum-dot transition through the resonance, using the applied bias voltage, while the excitation laser is kept at a constant wavelength. The measurement presented in Fig.\ \ref{fig:RF}(a) is carried out with a low excitation power corresponding to $22.5\,\%$ of the saturation count rate. In the best case, linewidths below $2\,{\mu}\text{eV}$ (full-width-at-half-maximum, FWHM) are measured on second time scales. This performance is comparable to that of quantum dots in thick diode structures located far from the GaAs-air interface\cite{Kuhlmann2013,Kuhlmann2014}. Narrow linewidths are reproducibly observed for different quantum dots in both samples (Fig.\ \ref{fig:RF}(b)). Quantum dot linewidths are strongly influenced by charge noise. This measurement demonstrates forcibly that the level of charge noise in the close-to-surface, p-i-n-i-n device is similar to the ultra-low charge noise in the very best far-from-surface, p-i-n device.  

\section{Electron spin pumping}
\label{sec:Electron spin pumping}
\begin{figure}[t!]
\includegraphics[width=0.9\columnwidth]{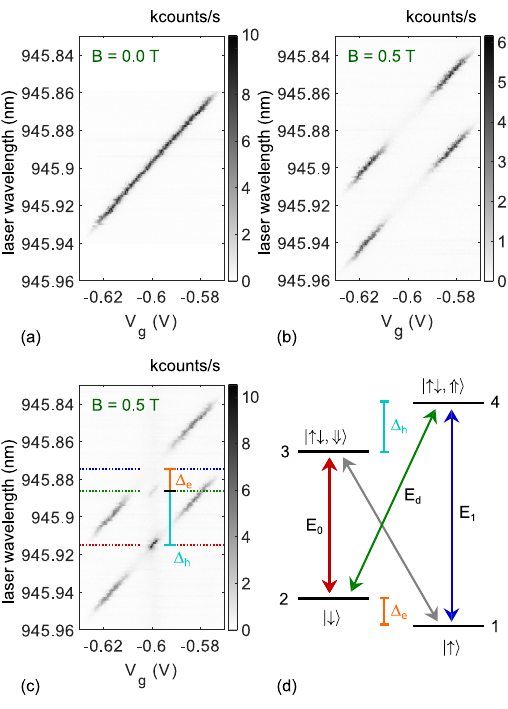}
\caption{(a) Resonance fluorescence of the singly charged exciton as a function of bias voltage and resonant laser wavelength. The measurement is carried out at zero external magnetic field on a quantum dot in sample 2. (b) RF of the same quantum dot at a magnetic field of $B=0.5\,\text{T}$ in the Faraday geometry. At the center of the plateau, the RF signal disappears due to optical spin pumping. (c) Resonant excitation is carried out with two lasers exciting the same quantum dot. The wavelength of the first laser is changed step-wise whereas the second laser is held at a constant wavelength of $945.874\,\text{nm}$ (indicated by the blue line). The signal reappears when both ``vertical" exciton transitions are excited simultaneously confirming the presence of optical spin pumping (indicated by the red line). When the second laser is in resonance with the ``diagonal" transition $\mid\downarrow\rangle\leftrightarrow\mid\uparrow\downarrow,\Uparrow\rangle$ (indicated by the green line), the RF signal is also enhanced since the second laser pumps the quantum dot back to its bright transition $\mid\uparrow\rangle\rightarrow\mid\uparrow\downarrow\text{,}\Uparrow\rangle$. However, this enhancement is weaker, since the corresponding transition is dipole ``forbidden", i.e.\ only weakly allowed. The observation of the ``diagonal" transition allows the Zeeman splittings, $\Delta_{\text{e}}$, $\Delta_{\text{h}}$ for electron and hole, respectively, to be determined. Note that in (a--c) all lasers are kept at the same power. (d) Level scheme of the quantum dot in the Faraday geometry.}
\label{fig:spin}
\end{figure}

Next we investigate the spin properties of a quantum dot by optical spin pumping experiments in a magnetic field in the Faraday geometry. The laser wavelength is changed stepwise to map the full Coulomb plateau. The background suppression of the dark-field microscope has a chromatic dependence and is therefore readjusted for each wavelength. In practice, this is carried out by an automatic algorithm that minimizes the intensity of the laser background by adjusting the polarization optics \cite{Kuhlmann2013a}. For a fixed laser wavelength the bias voltage is swept, sweeping the quantum dot transition with respect to the laser. This gives a ``horizontal" cut through the X$^{1-}$ exciton response, see Fig.\ \ref{fig:spin}(a). This procedure is repeated for different laser wavelengths giving a full map of the response over the single-electron Coulomb plateau. The results for zero magnetic field and a magnetic field of $0.5\,\text{T}$ (Faraday geometry) are shown in Fig.\ \ref{fig:spin}(a) and Fig.\ \ref{fig:spin}(b), respectively. Both measurements are done with the same excitation power. In Fig.\ \ref{fig:spin}(b) the $\text{X}^{1-}$-plateau shows a clear Zeeman splitting. Furthermore, the RF signal disappears in the middle of the plateau. This is the signature of optical spin pumping \citep{Atature2006,Dreiser2008,Kroner2008PRB}: the spin is initialized in one of the spin eigenstates. 

Spin pumping is interpreted in terms of the level scheme shown in Fig.\ \ref{fig:spin}(d). There are two strong transitions, the ``vertical" transitions, and two weak transitions, the ``diagonal" transitions. In the Faraday geometry, spin pumping arises due to the weakly allowed ``diagonal" transitions in combination with a long electron spin relaxation time. On driving the $\mid\uparrow\rangle\leftrightarrow\mid\uparrow\downarrow,\Uparrow\rangle$ transition, the electron is pumped into the $\mid\downarrow\rangle$ state via the weak ``diagonal" transition $\mid\uparrow\downarrow,\Uparrow\rangle\leftrightarrow\mid\downarrow\rangle$ (green line in Fig.\ \ref{fig:spin}(d)). The laser is no longer scattered by the quantum dot and the resonance fluorescence turns off. In the plateau center, the signal is reduced by a factor of $\alpha_{r}=40.1\pm1.6$ for the ``red" transition, and by a factor of $\alpha_{b}=37.6\pm1.2$ for the ``blue" transition, in both cases taking the RF intensity at zero magnetic field as a reference. To quantify the spin initialization we estimate a spin initialization fidelity $F=\sqrt{\langle\uparrow\mid\rho\mid\uparrow\rangle}$ for pumping the red, and $F=\sqrt{\langle\downarrow\mid\rho\mid\downarrow\rangle}$ for the blue transition. The initialization fidelity can be related to the resonance fluorescence via $F=\sqrt{1-1/\alpha_{r/b}}$ (see \hyperref[sec:appendix d]{appendix D} for details). This way we estimate initialization fidelities of $F=98.7\,\%$ for both spins. A significant difference is not expected at $4.2\,\text{K}$ and small magnetic fields as the thermal energy is much larger than the Zeeman splitting between the electron spin states. At the edges of the one-electron Coulomb plateau, the RF signal does not disappear. At the plateau edges, co-tunneling with the Fermi sea in the back gate randomizes the spin rapidly and spin pumping becomes ineffective \cite{Smith2005}. The observation of optical spin pumping in the Faraday geometry confirms that the spin-flip processes which couple the two electron spin states $\mid\uparrow\rangle$, $\mid\downarrow\rangle$ are significantly slower than the decay rate of the weakly allowed diagonal transition \cite{Dreiser2008}.

To confirm that the observed disappearance of the signal arises due to optical spin pumping, we repeat the experiment with a second laser, a re-pump laser \citep{Atature2006}. The second laser has a fixed wavelength of $945.87\,\text{nm}$, the wavelength of the ``vertical" transition $\mid\uparrow\rangle\leftrightarrow\mid\uparrow\downarrow,\Uparrow\rangle$ (blue arrow in Fig.\ \ref{fig:spin}(d)). These measurements are shown in Fig.\ \ref{fig:spin}(c). The laser powers are kept constant throughout the entire scan. Two re-pump resonances are observed (marked by red and green dashed lines in Fig.\ \ref{fig:spin}(d)). 

When the first laser, the ``pump" laser, is in resonance with the ``vertical" transition $\mid\downarrow\rangle\leftrightarrow\mid\uparrow\downarrow,\Downarrow\rangle$, the electron spin is shelved in the $\mid\uparrow\rangle$ state and with this laser alone, the RF disappears. However, in the presence of the re-pump laser, the electron spin is driven back into the $\mid\downarrow\rangle$ state and the RF reappears: the electron spin ends up in a statistical mixture of the two spin states. Similarly, the system ends up in a mixture of the spin states when the pump laser is stepped into resonance with the weakly-allowed ``diagonal" transition $\mid\downarrow\rangle\leftrightarrow\mid\uparrow\downarrow,\Uparrow\rangle$. However, since the ``diagonal" transition is only weakly allowed, the RF is relatively weak in this case. These observations explain the origin and intensity of the two re-pump resonances. 

The fact that the diagonal transition $\mid\downarrow\rangle\rightarrow\mid\uparrow\downarrow,\Uparrow\rangle$ is visible allows the energies of all three optical transitions to be determined. The energies of the different exciton transitions are denoted as $E_1$ for the transition $\mid\uparrow\rangle\rightarrow\mid\uparrow\downarrow,\Uparrow\rangle$, $E_0$ for $\mid\downarrow\rangle\rightarrow\mid\uparrow\downarrow,\Downarrow\rangle$, and $E_{\text{d}}$ for $\mid\downarrow\rangle\rightarrow\mid\uparrow\downarrow,\Uparrow\rangle$ (see Fig.\ \ref{fig:spin}(d)). The electron and hole Zemann splitting are given by $\Delta_{\text{e}}=E_1-E_{\text{d}}$ and $\Delta_{\text{h}}=E_{\text{d}}-E_0$. This allows the magnitude of the electron and hole g-factors to be determined via the relations $\Delta_{\text{e/h}}=g_{\text{e/h}}\mu_{\text{B}} B$. Assuming that the electron g-factor is negative, we find an electron g-factor of $g_{\text{e}}=-0.55$ and a hole g-factor of $g_{\text{h}} = 1.37$, values comparable to those in the literature \cite{Atature2006,Kroner2008PRL,Pinotsi2011}.

\section{Conclusions}
\label{sec:conclusion}
In conclusion, we have designed a p-i-n-i-n diode structure with a thickness of just $162.5\,\text{nm}$. The device enables single electron charging of embedded self-assembled quantum dots at low bias voltage and with small tunneling currents. The diode is fully compatible with the fabrication of photonic and phononic nanostructures in thin membranes. We demonstrate narrow optical linewidths and optical spin pumping for the close-to-surface quantum dots in the p-i-n-i-n diode. These excellent properties will underpin future exploitations of quantum dot spins in functionalized nanostructures.

\section{Acknowledgements}
\label{sec:acknowledgement}
MCL, IS, AVK and RJW acknowledge financial support from NCCR QSIT and from SNF Project No.\ 200020\_156637. This project has received funding from the European Union’s Horizon 2020 research and innovation programme under the Marie Skłodowska-Curie grant agreement No. 747866 (EPPIC). AJ, TP, LM, SS, and PL gratefully acknowledge financial support from the European Research Council (ERC Advanced Grant ``SCALE"), Innovation Fund Denmark (Quantum Innovation Center ``Qubiz"), and the Danish Council for Independent Research. SS acknowledges Villum Fonden for financial support. RS, AL, ADW gratefully acknowledge financial support from the grants DFH/UFA CDFA05-06, DFG TRR160, and BMBF Q.com-H 16KIS0109.

\section*{appendix a: analytical bandstructure model}
\label{sec:appendix a}
\begin{figure}[b!]
\includegraphics[width=0.9\columnwidth]{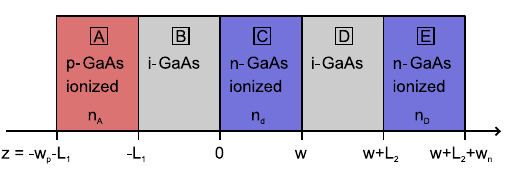}
\caption{Schematic p-i-n-i-n diode with labels used in the analytical calculation of the band structure. The letters (A--E) in the frames correspond to the different regions considered in the band structure calculation; the colors indicate the corresponding layers of the diode shown in Fig.\ \ref{fig:design}(c).}
\label{fig:for_analytic_calc}
\end{figure}
We present an analytic calculation of the band structure of the p-i-n-i-n diode \cite{Hosseini2015}. To this end, we divide the heterostructure in 5 different regions (Fig.\ \ref{fig:for_analytic_calc}). The first region (A) is the depletion zone of the p-type top gate, of width $w_p$ and doping concentration $n_A$, part of layer number 2 in Fig.\ \ref{fig:design}(c). The second region (B) is the intrinsic GaAs layer between top gate and an intermediate n-layer (layer 3 in Fig.\ \ref{fig:design}(c)). Its width is denoted by $L_1$. The third region (C) is the intermediate n-layer (layer 4 in Fig.\ \ref{fig:design}(c)) with a width denoted by $w$ and doping concentration $n_d$. The fourth region (D) is the subsequent intrinsic region of width $L_2$ which includes the quantum dot layer (layer 5, 6 in Fig.\ \ref{fig:design}(c)). The final region (E) is the depletion zone of the back gate (part of layer 7 in Fig.\ \ref{fig:design}(c)). It has a width of $w_n$ and a doping concentration of $n_D$.

We apply the Poisson equation $\Delta\Phi=-\frac{e\cdot n}{\epsilon\epsilon_0}$ to all five regions ($e$ electron charge, $\epsilon_0$ vacuum permittivity, $\epsilon$ relative permittivity of GaAs, and $n$ carrier density). Note that the potential $\Phi$ is defined for a positive probe charge and has to be reversed in sign to describe an electron in the conduction band. Together with the constraints that the electric displacement field $-\epsilon\epsilon_0\cdot\frac{\partial\Phi}{\partial z}$ must be continuous and vanishes at the outer edges of the depletion zones, one obtains the following 5 equations for the electric field in the different regions A--E of the structure:
\begin{align}
&\text{A}:\,\frac{\partial\Phi}{\partial z}=\frac{e}{\epsilon\epsilon_0}\cdot n_A\cdot\left(z+w_p+L_1\right)\label{Efield_Equ1}\\
&\text{B}:\,\frac{\partial\Phi}{\partial z}=\frac{e}{\epsilon\epsilon_0}\cdot n_A w_p\label{Efield_Equ2}\\
&\text{C}:\,\frac{\partial\Phi}{\partial z}=\frac{e}{\epsilon\epsilon_0}\cdot \left(n_A w_p-n_d z\right)\label{Efield_Equ3}\\
&\text{D}:\,\frac{\partial\Phi}{\partial z}=\frac{e}{\epsilon\epsilon_0}\cdot \left(n_A w_p-n_d w\right)\label{Efield_Equ4}\\
&\text{E}:\,\frac{\partial\Phi}{\partial z}=\frac{e}{\epsilon\epsilon_0}\cdot \left( n_D\cdot\left(L_2+w-z\right) +n_A w_p - n_d w\right)\label{Efield_Equ5}
\end{align}
Integration of the electric field in all 5 regions of the diode yields the potential drop $\Delta V$ between top gate and back gate:
\begin{align}
\frac{\epsilon\epsilon_0}{e}\cdot\Delta V = &\frac{\epsilon\epsilon_0}{e}\cdot\left(V_{\text{0}}-V_{\text{bias}}\right)\nonumber\\
= &\frac{n_A}{2} w_p^2 + n_A w_p L_1 + n_A w_p w - \frac{n_d}{2}w^2\nonumber\\
&+L_2\cdot\left(n_A w_p - n_d w\right) -\frac{n_D}{2} w_n^2\nonumber\\
&+w_n\cdot\left(n_A w_p-n_d w\right)\label{potential_analytic}
\end{align}
where $V_{\text{0}}$ is the built-in voltage of the diode and $V_{\text{bias}}$ is the externally applied bias voltage. For high doping concentrations when top and back gate are degenerately doped, the built-in voltage is given by: $e\cdot V_{\text{0}}=E_{\text{gap}}+E_{\text{F}}^{\text{e}}+E_{\text{F}}^{\text{h}}$ where $E_{\text{gap}}$ is the band gap of GaAs and $E_{\text{F}}^{\text{e/h}}$ is the Fermi level for electrons in the back gate and holes in the top gate, respectively ($E_{\text{F}}^{\text{e/h}}=\hbar^2/2m_{e/h}^{\text{*}}\cdot\left(3\pi^2 n \right)^{2/3}$). The condition that the entire device is charge neutral,
\begin{align}\label{charge_neutral}
-n_A\cdot w_p+n_d\cdot w+n_D\cdot w_n=0,
\end{align}
in combination with Eq.\ \ref{potential_analytic}, determines the widths of the depletion zones $w_p$ and $w_n$:
\begin{align}\label{wp_width}
w_p= &\frac{1}{a_1}\cdot\left(a_2+\sqrt{a_2^2+2 a_1 a_3}\right)\nonumber\\
&a_1 = n_A+\frac{n_A^2}{n_D}\nonumber\\
&a_2 = -n_AL_2 -n_Aw - n_AL_1 + \frac{n_A n_d}{n_D} w\nonumber\\
&a_3 = n_dL_2w +\frac{n_dw^2}{2}+ \frac{\epsilon\epsilon_0}{e}\Delta V-\frac{n_d^2w^2}{2 n_D}\nonumber\\
w_n=&\frac{1}{n_D}\cdot\left(n_Aw_p-n_dw\right)
\end{align}
Using Eq.\ \ref{wp_width} the potential as a function of vertical position inside the heterostructure is obtained by integration over Eq.\ \ref{Efield_Equ1}--\ref{Efield_Equ5}. In particular, the electric field at the position of a quantum dot is given by Eq.\ \ref{Efield_Equ4}.

\section*{appendix b: bias voltage of Coulomb plateaus}
\label{sec:appendix b}
\begin{figure}[t!]
\includegraphics[width=0.9\columnwidth]{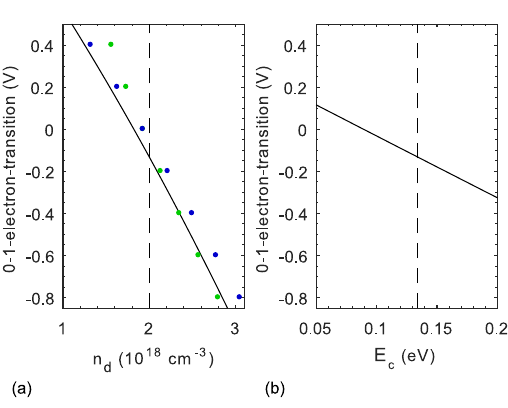}
\caption{(a) Calculated shift of the 0-1 electron charging transition as a function of the doping of the intermediate n-type layer $n_d$. The back gate doping is scaled correspondingly ($n_d=n_D$). The black curve shows the result of the analytical calculation (see \hyperref[sec:appendix a]{appendix A}); the green points represent the results of band structure simulations including surface depletion; the blue points represent the results of band structure simulations assuming ohmic contacts. (b) Shift of the 0-1 electron charging transition as a function of the single electron confinement energy $E_c$.}
\label{fig:error_ndope}
\end{figure}

We present a possible explanation for the fact that the 0-1 electron transition takes place at a bias voltage of $V_{\text{bias}}=-0.6\,\text{V}$ and not around zero bias as intended. Deviations of heterostructure or quantum dot parameters can shift this transition voltage. The part of the heterostructure that influences the 0-1e transition voltage most strongly is the intermediate n-type layer. A deviation in its thickness or its doping concentration can change the electric field experienced by the quantum dot. The layer thickness can be controlled rather precisely in MBE-growth and we thus simulate the 0-1e transition voltage as a function of the doping concentration of the intermediate n-layer. The ratio between doping of the intermediate n-layer and doping of the backgate is kept constant for this estimation since a systematically different n-doping would affect both layers. We use the analytical model and also numerical band structure simulations (nextnano). We take a single electron confinement energy of the quantum dot of $E_{\text{c}}=134\,\text{meV}$ \cite{Seidl2005} and vary the doping concentration (Fig.\ \ref{fig:error_ndope}(a)). The 0-1 electron transition voltage obtained numerically assuming ohmic boundary conditions agrees well with the analytical model, but is systematically slightly larger. We explain this by the fact that the numerical Poisson equation solver takes into account a charge overspill of back gate electrons into the intrinsic region (see Fig.\ \ref{fig:design}(c)). This effect lifts the conduction band energy slightly at the location of the QDs. In contrast the analytical model assumes abrupt depletion regions. A numerical simulation taking into account surface depletion via Schottky barriers of $1\,\text{V}$ gives comparable results (see Fig.\ \ref{fig:error_ndope}). Surface effects are not considered in the analytical model. All this work predicts a 0-1e transition voltage of about
$-0.1\,\text{V}$ for the nominal doping concentration $n_d=2.0\cdot10^{18}\,\text{cm}^{-3}$.

Fig.\ \ref{fig:error_ndope}(b) shows the 0-1 electron transition voltage as a function of the quantum dot single electron confinement energy $E_{\text{c}}$ keeping the doping at the nominal value of $n_d=2.0\cdot10^{18}\,\text{cm}^{-3}$. The dashed black line indicates a single electron confinement energy of $E_c=134\,\text{meV}$ that has been reported in literature \citep{Seidl2005}. One can see that the shift of the 0-1e transition voltage to $-0.6\,\text{V}$ cannot be explained by any realistic single electron confinement potential of the quantum dot. This suggests that the most likely explanation for the shift of the 0-1e transition to $-0.6\,\text{V}$ is a deviation of the n-doping from the nominal value. An increase by about $30\,\%$ reproduces the experimental result taking  $E_{\text{c}}=134\,\text{meV}$ (Fig.\ \ref{fig:error_ndope}(a)). A reduced doping of the p-type top gate would also shift the 0-1e transition to more negative bias voltages. However, the effect of an under-doped p-layer is smaller and cannot explain the shift to $-0.6\,\text{V}$ completely.

\section*{appendix c: supplementary experimental data}
\label{sec:appendix c}
\begin{figure}[t!]
\includegraphics[width=0.9\columnwidth]{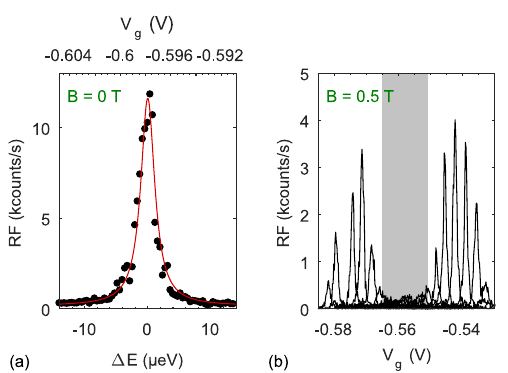}
\caption{(a) Resonance fluorescence of a quantum dot in sample 2 (the one from Fig.\ \ref{fig:spin}). The linewidth obtained by a Lorentzian fit (red curve) is $2.7\,{\mu}\text{eV}$ for a power corresponding to $27\,\%$ of the saturation count rate. (b) Optical spin pumping on the singly charged exciton for a quantum dot in sample 1 (the one from Fig.\ \ref{fig:RF}) at a magnetic field of $0.5\,\text{T}$. The different peaks correspond to RF measurements for different excitation wavelengths. The gray shaded region indicates the regime where the spin pumping dominates over co-tunneling processes and the RF signal thus disappears. At the plateau edges, the co-tunneling dominates and resonance fluorescence reappears.}
\label{fig:additional}
\end{figure}

In the main text, we presented narrow linewidths for a quantum dot in sample 1 whereas optical spin pumping is demonstrated for a quantum dot in sample 2. To illustrate that our measurements are reproducible on different quantum dots, we show in Fig.\ \ref{fig:additional} a typical linewidth for the quantum dot in sample 2, and demonstrate optical spin pumping for the quantum dot in sample 1.

\section*{appendix d: analysis of spin pumping}
\label{sec:appendix d}
Here we show how resonance fluorescence of the singly charged exciton is used to obtain spin initialization fidelities. In section \ref{sec:Electron spin pumping}, the initialization fidelity was connected to the ratio $\alpha_{r/b}$ between the RF-intensity when no spin pumping is present (at $\text{B}=0\,\text{T}$) and the RF-intensity when spin pumping is active (at $\text{B}=0.5\,\text{T}$).

We give a derivation of this relation in a rate-equation picture. At zero magnetic field, both allowed transitions $\mid\downarrow\rangle\leftrightarrow\mid\uparrow\downarrow,\Downarrow\rangle$ and $\mid\uparrow\rangle\leftrightarrow\mid\uparrow\downarrow,\Uparrow\rangle$ are degenerate (see Fig.\ \ref{fig:spin}(d)). In the steady state, the ratio between occupation of upper and lower levels is given by:
\begin{equation}
\label{A_steady_state}
\frac{N_3^{B=0}}{N_2^{B=0}}=\frac{N_4^{B=0}}{N_1^{B=0}}=\frac{\Gamma}{\Gamma+\gamma+\gamma_D}\equiv b
\end{equation}
with $N_3^{B=0}=N_4^{B=0}$ the occupation of the excited states and $N_2^{B=0}=N_1^{B=0}$ the occupation of the ground states (see Fig.\ \ref{fig:spin}(d) for labels). $\Gamma$ denotes the stimulated emission/excitation rate, $\gamma$ the spontaneous emission rate via the dipole-allowed ``vertical" transitions, and $\gamma_D$ the spontaneous emission rate via the ``diagonal" transitions. The resonance fluorescence intensity $\text{RF}^{B=0}$ is directly connected to the occupation of the upper states: $\text{RF}^{B=0}=\tilde{c}\left(N_3^{B=0}+N_4^{B=0}\right)$. The combination of Eq.\ \ref{A_steady_state} and the normalization condition $\sum_{i=1}^4 N_i^{B=0}=1$ yields the equation:
\begin{equation}
\label{A_RF}
\text{RF}^{B=0}=\tilde{c}\left(N_3^{B=0}+N_4^{B=0}\right)=\frac{\tilde{c}}{1+1/b}.
\end{equation}

In finite magnetic field, the transitions $\mid\downarrow\rangle\leftrightarrow\mid\uparrow\downarrow,\Downarrow\rangle$ and $\mid\uparrow\rangle\leftrightarrow\mid\uparrow\downarrow,\Uparrow\rangle$ are split in energy. We take the case when the red-shifted transition $\mid\downarrow\rangle\leftrightarrow\mid\uparrow\downarrow,\Downarrow\rangle$ is driven by a laser field whereas the other one is not addressed. This means that $N_4^{B\neq0}=0$ and the resonance fluorescence is connected to the occupation of just one upper level: $\text{RF}^{B\neq0}=\tilde{c}N_3^{B\neq0}$. In the steady state, the ratio of $N_3^{B\neq0}$ and $N_2^{B\neq0}$ is also given by the relation $N_3^{B\neq0}/N_2^{B\neq0}=b$, see Eq.\ \ref{A_steady_state}. In combination with the normalization condition $\sum_{i=1}^3 N_i^{B\neq0}=1$ this leads to:
\begin{equation}
\label{B_RF}
\text{RF}^{B\neq0}=\tilde{c}N_3^{B\neq0}=\tilde{c}\frac{1-N_1^{B\neq0}}{1+1/b}.
\end{equation}
The combination of Eq.\ \ref{A_RF} and Eq.\ \ref{B_RF} directly connects the occupation $N_1^{B\neq0}$ of the ground state with the resonance fluorescence intensities:
\begin{equation}
\label{eq:N3}
N_1^{B\neq0}=1-\frac{1}{\alpha_r}\text{, with }\alpha_r=\frac{\text{RF}^{B=0}}{\text{RF}^{B\neq0}}.
\end{equation}
This equation shows how the initialization fidelity can be deduced from the RF-measurements.

\begin{figure}[t!]
\includegraphics[width=0.9\columnwidth]{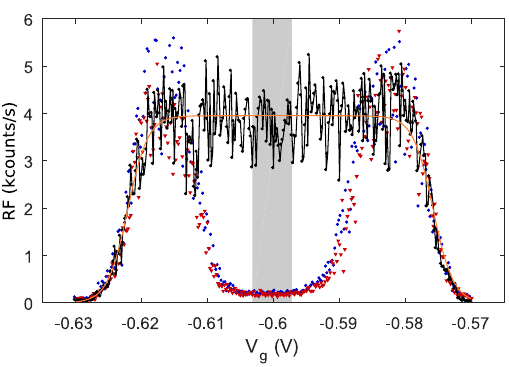}
\caption{Resonance fluorescence intensity along the single electron Coulomb plateau for the quantum dot shown in Fig.\ \ref{fig:spin}. The black curve shows data at $0.0\,\text{T}$ divided by a factor of two (to give a signal per spin); the red triangles (blue circles) show the data at $0.5\,\text{T}$ for the lower (higher) frequency Zeeman transitions. To obtain the ratio between the plateau intensities with and without spin pumping, the corresponding signals are averaged in the plateau center (gray shaded region). The orange curve is a fit of the data at $B=0.0\,\text{T}$ to Eq.\ \ref{2sidedFermi}.}
\label{spin_additional}
\end{figure}

The corresponding values of $\alpha_{r/b}$ are determined in the following way. For every bias voltage, the maximum RF signal (Fig.\ \ref{fig:spin}) is determined. Results are plotted in Fig.\ \ref{spin_additional} for the two transitions at $0.5\,\text{T}$ (Zeeman-split) and for the single transition at zero magnetic field. The signals are averaged over a small region in the Coulomb plateau center to determine accurately the strength of the resonance fluorescence in the regime where optical spin pumping dominates over spin co-tunneling. In this way, we determine the intensity ratio $\alpha_{r/b}$ between the signal at zero magnetic field and the signal at $B=0.5\,\text{T}$.

\section*{appendix e: lever arm approximation}
\label{sec:appendix e}
On increasing the bias voltage from the center of the Coulomb plateau, the X$^{1-}$ RF drops once it becomes electrically favorable for a second electron to tunnel into the quantum dot. On the other hand, on decreasing the bias voltage from the plateau center, the X$^{1-}$ RF drops once it becomes energetically favorable for the electron to tunnel out of the quantum dot. In both cases, the edges of the X$^{1-}$ plateau are not abrupt since the electron occupation in the back gate is determined by a thermally smeared Fermi distribution. At its edges, the X$^{1-}$ RF signal maps the Fermi distribution of the back gate and is well described by a 2-sided Fermi-distribution:
\begin{equation}
\label{2sidedFermi}
I_{\text{RF}}(V)=I_0\cdot\frac{1}{1+\exp\left(\frac{e\cdot\left(V-V_1\right)}{\lambda_{\text{diff}}k_{\text{B}}T}\right)} \cdot \frac{1}{1+\exp\left(\frac{e\cdot\left(V_2-V\right)}{\lambda_{\text{diff}}k_{\text{B}}T}\right)}
\end{equation}
where $k_{\text{B}}T$ is the thermal energy, $V_1$ and $V_2$ specify the bias voltage at the plateau edges, and $I_0$ is the intensity in the plateau center. The variable $\lambda_{\text{diff}}$, the differential lever arm, is defined by $\lambda_{\text{diff}}=e\cdot\left(\frac{\partial\Phi_{\text{QD}}}{\partial V_{\text{bias}}}\right)^{-1}$ where $\Phi_{\text{QD}}$ is the energy difference between back gate Fermi-energy and the quantum dot single electron level. Thus, $\lambda_{\text{diff}}$ parameterizes how the potential of the quantum dot changes with bias voltage $V_{\text{bias}}$.

We determined the differential lever arm as a function of the bias voltage by using numerical band structure simulations. We find a value of $\lambda_{\text{diff}}=4.17$ at a bias voltage of $-0.6\,\text{V}$. A slightly increased n-doping explaining the 0-1e transition at this bias is taken into account (see \hyperref[sec:appendix b]{appendix B}). In the simulation, the lever arm is to a good approximation constant over the single electron Coulomb plateau. We fit the model described by Eq.\ \ref{2sidedFermi} using the position of the plateau edges $V_1$, $V_2$ and the plateau intensity $I_0$ as the only fit parameters. The temperature $4.2\,\text{K}$ as well as the differential lever arm are fixed parameters in the fit. The fit describes the experimental data very well (Fig.\ \ref{spin_additional}). This is further evidence that the electrical properties of our sample are well understood.

As a final remark we note that often the lever arm is also defined as $\lambda_{\text{geo}}=L/L_{QD}$ (geometrical lever arm) and $\lambda_{\text{el}}=e\cdot\left(\frac{\Phi_{\text{QD}}}{V_{\text{0}}-V_{\text{bias}}}\right)^{-1}$ (electrical lever arm). For diode structures with little band bending, the three parameters $\lambda_{\text{diff}}$, $\lambda_{\text{el}}$, and $\lambda_{\text{geo}}$ are to a good approximation equivalent \citep{Smith2005,Dreiser2008,Lei2008}. Obviously, this is not the case for the heterostructure presented here as a result of band bending in the p-i-n-i-n structure.

\bibliography{main.bbl}

\end{document}